\documentclass[5p,authoryear,times]{elsarticle}

\usepackage{amsmath}
\usepackage{amssymb}
\usepackage{txfonts}
\usepackage{epsfig,graphicx}
\usepackage{color}
\usepackage[colorlinks=true, linkcolor=blue, citecolor=blue, urlcolor=blue]{hyperref}
\bibpunct{(}{)}{;}{a}{}{,} 

\journal{Astronomy and Computing}

\newcommand{\txd}{{\text{d}}}
\newcommand{\txs}{{\text{s}}}

\newcommand{\bfa}{{\boldsymbol{a}}}

\newcommand{\bfx}{{\boldsymbol{x}}}
\newcommand{\bfy}{{\boldsymbol{y}}}

\newcommand{\aap}{Astron. Astroph.\ }
\newcommand{\aj}{Astronom. J.\ }
\newcommand{\apj}{Astrophys. J.\ }

\newcommand{\apjs}{Astrophys. J. Suppl.\ }
\newcommand{\araa}{Ann. Rev. Astron. Astroph.\ }
\newcommand{\mnras}{Mon. Not. R. Astron. Soc.\ }

\begin{document}

\begin{frontmatter}

\title{SKIRT: the design of a suite of input models \\for Monte Carlo radiative transfer simulations}

\author{M. Baes and P. Camps}
\address{Sterrenkundig Observatorium, Universiteit Gent, Krijgslaan 281 S9, B-9000 Gent, Belgium}




\begin{abstract}
The Monte Carlo method is the most popular technique to perform radiative transfer simulations in a general 3D geometry. The algorithms behind and acceleration techniques for Monte Carlo radiative transfer are discussed extensively in the literature, and many different Monte Carlo codes are publicly available. On the contrary, the design of a suite of components that can be used for the distribution of sources and sinks in radiative transfer codes has received very little attention. The availability of such models, with different degrees of complexity, has many benefits. For example, they can serve as toy models to test new physical ingredients, or as parameterised models for inverse radiative transfer fitting. For 3D Monte Carlo codes, this requires algorithms to efficiently generate random positions from 3D density distributions. 

We describe the design of a flexible suite of components for the Monte Carlo radiative transfer code SKIRT. The design is based on a combination of basic building blocks (which can be either analytical toy models or numerical models defined on grids or a set of particles) and the extensive use of decorators that combine and alter these building blocks to more complex structures. For a number of decorators, e.g.\ those that add spiral structure or clumpiness, we provide a detailed description of the algorithms that can be used to generate random positions. Advantages of this decorator-based design include code transparency, the avoidance of code duplication, and an increase in code maintainability. Moreover, since decorators can be chained without problems, very complex models can easily be constructed out of simple building blocks. Finally, based on a number of test simulations, we demonstrate that our design using customised random position generators is superior to a simpler design based on a generic black-box random position generator.
\end{abstract}

\begin{keyword}
radiative transfer --- methods: numerical --- designing software --- design patterns\vspace*{1ex}
\end{keyword}

\end{frontmatter}

\section{Introduction}

\noindent
Radiative transfer, the process that describes radiation propagating through and interacting with matter, is a general problem that is encountered in all areas of astronomy (and far beyond). Due to its high dimensionality and nonlocal and nonlinear behaviour, it is generally considered as one of the most challenging problems in numerical astrophysics.

Recent years have seen an impressive advancement of three-dimensional (3D) radiative transfer studies, thanks to the increase in computational power, the availability of a wealth of new observational constraints and the development of new algorithms. Among the different approaches available to solve the radiative transfer problem, the Monte Carlo method is generally the most popular one. The first Monte Carlo radiative transfer codes were developed more than four decades ago \citep[e.g.,][]{1970A&A.....9...53M, 1974ApJ...190...67R, 1977ApJS...35....1W}, and since then the method has steadily increased its market share in all fields where 3D radiative transfer is important. Many powerful Monte Carlo codes are available in various fields of computational astrophysics, including dust radiative transfer \citep[][and references therein]{2013ARA&A..51...63S}, Ly$\alpha$ line transfer \citep{2006A&A...460..397V, 2006ApJ...649...14D, 2009ApJ...704.1640L}, ionising radiation transport \citep{2001MNRAS.324..381C, 2003MNRAS.345..379M, 2004MNRAS.348.1337W}, neutron transport \citep{Romano2013274} and neutrino radiation transport \citep{2012ApJ...755..111A}. These and many other papers and monographs discuss at length the main algorithms behind Monte Carlo radiative transfer and the various proposed improvements and acceleration techniques. 

A very different aspect of Monte Carlo radiative transfer codes that has received very little attention in the literature, is the setup of a suite of components that describe the distribution of the sources and sinks in the radiative transfer code (i.e., the objects that add radiation to or remove radiation from the radiation field). This is an aspect outside the real core of Monte Carlo radiative transfer problem. Virtually all of the available codes require the user to hard-code the model makeup for each distinct problem (although the results of hydrodynamical simulations can generally be processed out of the box, given the appropriate import module). We argue, however, that it is very useful for a radiative transfer code to provide a suite of input models, or \emph{geometries}, as built-in components. Such toy models can provide a low-threshold introduction to new users. They also have a crucial role in a more scientific way: toy input models are used to benchmark different codes \citep{1997MNRAS.291..121I, 2004A&A...417..793P, 2009A&A...498..967P}, to investigate the effects of dust attenuation on the apparent structural parameters of galaxies \citep{2006A&A...456..941M, 2010MNRAS.403.2053G, 2013A&A...553A..80P, 2013A&A...557A.137P}, or to investigate the physical impact of e.g.\ clumpiness or spiral arms \citep{1996ApJ...463..681W, 2000MNRAS.311..601B, 2000A&A...353..117M, 2006BaltA..15..581S}. Finally, they are essential for so-called inverse radiative transfer, i.e.\ when a parameterised radiative transfer model is fitted to observational data \citep{1999A&A...344..868X, 2007A&A...471..765B, 2012ApJ...746...70S, 2013A&A...550A..74D, 2014MNRAS.441..869D}. 

In order to be useful for these goals, the suite of input models should be diverse enough and contain models with different degrees of complexity, ranging from the metaphorical spherical cows to more realistic toy models that consider, for example, spheroidal and triaxial geometries, including bars, spiral arms and clumpy distributions. Setting up such a suite is more complex than it might appear at first sight. While each model is in principle completely determined by the 3D density distribution $\rho(\bfx)$, the setup requires more than just an implementation for this density function. 

A crucial aspect of Monte Carlo radiative transfer codes is the emission of a multitude of simulated photon packets from random locations sampled from the source density distribution. So each model in the suite should contain a routine that generates random numbers according to $\rho(\bfx)$. Moreover, this random position generation needs to be very efficient, since the random positions are sampled extremely often. In this aspect, Monte Carlo radiative transfer simulations differ from other codes where random positions need to be sampled from arbitrary density distributions, e.g.\ to generate the initial conditions for N-body or hydrodynamical simulations. The efficient generation of random positions from complex 3D density distribution is not a trivial task.

Another aspect that needs consideration is the organisation of such a suite of input models. One could provide a parameterised model that can cover every possible option, but this would quickly lead to an explosion of options that is hard to overview and maintain, and would inevitably contain substantial code duplication. 

In this paper, we describe how these issues are dealt with in the publicly available 3D Monte Carlo dust radiative transfer code SKIRT\footnote{SKIRT documentation: http://www.skirt.ugent.be\\ \indent SKIRT code repository: https://github.com/skirt/skirt} \citep{2003MNRAS.343.1081B, 2011ApJS..196...22B, 2015A&C.....9...20C}. While we summarise the relevant information for the discussion here, an in-depth description of the architecture and overall design of the SKIRT code can be found in \citet{2015A&C.....9...20C}. It should also be noted that the presented design issues, while discussed in the context of SKIRT and thus dust radiative transfer, are fully applicable to other Monte Carlo transport problems.

This paper is laid out as follows. In Section~{\ref{Techniques.sec}} we review the general techniques to generate random vectors from multivariate distributions that are available in the specialised numerical analysis literature, but less so in the astrophysics community. In Section~{\ref{GeometrySetup.sec}} we present the general setup of the suite of models for the SKIRT code. In Section~{\ref{BuildingBlocks.sec}} we describe the various building blocks currently present in the SKIRT code, and in Section~{\ref{Decorators.sec}} we present a number of decorators that can be used to combine and add complexity to simple building blocks. Some decorators are analysed in more detail, focusing on the methods used to efficiently generate random positions. In Section~{\ref{Discussion.sec}} we discuss the advantages of the decorator-style design of our suite of components, and we critically investigate an alternative option in which random positions are generated using a generic routine rather than customised generators. We show that, while such an approach is simpler to implement and might hence seem an attractive alternative, it cannot compete with our approach, in terms of accuracy and efficiency. Section~{\ref{Conclusions.sec}} sums up.

\section{Multivariate random number generation techniques}
\label{Techniques.sec}

\noindent
The generation of random numbers from univariate distributions is a well-known topic in numerical analysis. A number of standard methods are widely used and clearly described in standard textbooks \citep[e.g.,][]{Press2007}. Additional techniques that are less known in the (astro)physics community include the acceptance-complement method \citep{Kronmal1981,Kronmal1984} and the Forsythe-von Neumann method \citep{vonNeumann1951, Forsythe1972}. For an extensive overview of random number generation from univariate distribution, see \citet{Devroye1986}. Unfortunately, the generation of random vectors from multivariate distributions is much more complex. The only multivariate distributions from which random vectors can be generated directly are those where the density can be written as a product of independent univariate density distributions. In general, however, more advanced techniques are necessary.

\subsection{The inversion method}
\vspace{0.5em}

\noindent
The inversion method, also known as inversion sampling, is the most popular method for univariate generation problems. The basis of this method is the following: if $y$ is distributed according to a density\footnote{Throughout this paper, all densities, univariate or multivariate, are assumed to be properly normalised.} $g(y)$, then the variable $x$ defined as the solution of the equation $y = y(x)$ is distributed according to the density
\begin{equation}
f(x) = g\Bigl(y(x)\Bigr)\,\left|\,\frac{\txd y}{\txd x}(x)\,\right|
\label{transfmethod1D}
\end{equation}
If we now want to generate a number from a given density $f(x)$, we can take a uniform distribution \begin{equation}
g(y)=
\begin{cases}
\;1\quad&\text{if $0\leq y\leq 1$,} \\
\;0&\text{else,}
\end{cases}
\end{equation} 
and set $y(x) = F(x)$ where $F$ is the cumulative distribution corresponding to the density $f(x)$. The inversion method can hence be used to generate random variates with an arbitrary density, provided that the inverse function $F^{-1}(y)$ to the cumulative distribution $F(x)$ is explicitly known. Classical examples include the exponential distribution, the Cauchy distribution, the Rayleigh distribution and the logistic distribution.

Formula (\ref{transfmethod1D}) can directly be expanded to multiple dimensions: if $\bfy$ is distributed according to the joint probability distribution $g(\bfy)$, then the vector $\bfx$ defined as the solution of the vector equation $\bfy=\bfy(\bfx)$ is distributed according to the joint distribution
\begin{equation}
f(\bfx) = g\Bigl(\bfy(\bfx)\Bigr)\,\left|\,\frac{\partial\bfy}{\partial\bfx}(\bfx)\,\right|
\label{transfmethodmD}
\end{equation}
where $|\partial\bfy/\partial\bfx|$ is the Jacobian determinant. Probably the most famous application of this formula is the Box-Muller method to generate random normally distributed deviates \citep{Box1958, Bell1968}. Another interesting special application is the case of a linear transformation. In this case, the transformation $\bfy = \bfy(\bfx)$ can be written as a matrix multiplication $\bfy = {\bf{H}}\,\bfx$, and the distribution of $\bfx$ is 
\begin{equation}
f(\bfx) = |\,{\bf{H}}\,|\,g({\bf{H}}\,\bfx)
\label{transfmethodlin}
\end{equation}
with $|\,{\bf{H}}\,|$ the absolute value of the determinant of ${\bf{H}}$. This kind of transformation is particularly useful for the generation of random vectors with a given dependence structure, as measured by the covariance matrix \citep{Scheuer1962, Barr1972}. 

In general, however, it is not straightforward to use this identity (\ref{transfmethodmD}) to construct a method that can be used to generate random positions from an arbitrary multidimensional distribution. 

\subsection{The rejection method}
\vspace{0.5em}

\noindent
Apart from the inversion method, the rejection method, also known as the acceptance-rejection method, is the most popular method to sample nonuniform random numbers from univariate distributions \citep{vonNeumann1951, Devroye1986}. The basic idea behind the method is that, if one wants to sample a random number from a function $f(x)$, one can sample uniformly from the 2D region under the graph $f(x)$. More concretely, assume that $f(x)\leqslant c f_{\text{ref}}(x)$, where $f_{\text{ref}}(x)$ is another distribution from which random numbers are easily generated, and $c$ is the so-called rejection constant (it obviously satisfies the condition $c\geqslant1$). One then generates a uniform deviate $\xi$ and a random number $x$ from the reference distribution $f_{\text{ref}}(x)$, and calculates the quantity $t = \xi c f_{\text{ref}}(x)/f(x)$. This procedure is repeated until $t\leqslant1$, in which case $x$ is the desired random number.

One of the advantages of the rejection method is that it does not require that the cumulative distribution function be analytically known, let alone be invertible. However, its effectiveness depends on how accurate $f$ is approximated from above by $c f_{\text{ref}}$. Less accurate approximation leads to a greater chance of rejection; on average $c$ iterations of the loop are required before one successful random number is generated. Moreover, the reference distribution should be such that random numbers can be easily generated from it and that the computation of $c f_{\text{ref}}(x)/f(x)$ is simple. For a range of standard distributions, such as the gamma distribution and the Poisson distribution, efficient reference functions can be constructed. 

The rejection method is by no means limited to univariate distributions, and can immediately be applied to multivariate generation problems. However, the rejection rate typically increases rapidly when going from one to more dimensions, which decreases the efficiency of the method. Moreover, the design of a suitable reference  distribution becomes much more complicated.

\subsection{The composition method}
\vspace{0.5em}

\noindent
The composition method or probability mixing method is another important technique in both univariate and multivariate random number generation \citep{Marsaglia1964, Hormann2004}. Rather than a method on itself to generate non-uniform random numbers, it is a principle to facilitate and speed up other random number generating methods. The simple idea behind composition is to decompose the distribution $f(\bfx)$ as a weighted sum,
\begin{equation}
f(\bfx) = \sum_{i=1}^K w_i f_i(\bfx)
\end{equation}
with all $f_i$ normalised densities, and the weights $w_i$ a probability vector (i.e., all $w_i\geqslant0$ and $\sum_i w_i=1$). To generate a random $\bfx$ from the distribution $f(\bfx)$, we first generate a random integer number $k$ from the discrete probability vector $w_i$, and subsequently generate a random $\bfx$ from the density $f_k(\bfx)$. A prerequisite for this method to work is that, obviously, the decomposition can be done efficiently, and that the complexity of the problem is reduced by the decomposition. 

\subsection{The conditional distribution method}
\vspace{0.5em}

\noindent
Finally, a powerful technique that applies only to multivariate random number generation is the so-called conditional distribution method \citep{Devroye1986, Hormann2004}. It is based on the Bayesian identity
\begin{equation}
f(x_1,x_2) = f_1(x_1)\,f_2(x_2\,|\,x_1)
\end{equation}
which expresses that a joint distribution of two variables can be calculated as the product of the marginal distribution of the former variable and the conditional distribution of the latter. Expanding it to multiple dimensions, 
\begin{equation}
f(\bfx) = f_1(x_1)\,f_2(x_2\,|\,x_1)\cdots f_N(x_N\,|\,x_1,\ldots,x_{N-1})
\label{cdm}
\end{equation}
The beauty of this technique is that it reduces the multidimensional generation problem to a sequence of independent univariate generation problems. The main drawback is that it can only be used when much detailed information is known about the distribution $f(\bfx)$. In particular, it is by far not always the case that marginal and conditional distributions are easily calculated in closed form. The standard textbook example of this technique is the multivariate Cauchy distribution.

\section{General setup of the SKIRT {\tt{Geometry}} suite}
\label{GeometrySetup.sec}

\noindent
SKIRT is a multi-purpose 3D Monte Carlo dust radiative transfer code that is mainly used to simulate dusty galaxies and active galactic nuclei \citep[e.g.,][]{2002MNRAS.335..441B, 2010A&A...518L..39B, 2012MNRAS.420.2756S, 2012MNRAS.419..895D, 2014A&A...571A..69D}. The code was designed as a highly modular software, with a particular consideration for the development of a flexible and easy user interface and the use of proven software design principles as described by \citet{2015A&C.....9...20C}.

The SKIRT code offers a wealth of configurable features that are ready to use without any programming. In particular, the SKIRT code is equipped with a suite of input model components, the so-called {\tt{Geometry}} classes, that can be used to represent distributions of either sources or sinks. Essentially, the suite consists of a number of classes that all inherit from an abstract {\tt{Geometry}} class, for which the C++ class interface looks like

{
\small
\begin{verbatim}
class Geometry {
  public:
    Geometry();
    virtual ~Geometry();
    virtual double density(Position x) const = 0;
    virtual Position generatePosition() const = 0;
}
\end{verbatim}
}

\noindent Both the {\tt{density}} and {\tt{generatePosition}} functions are pure virtual functions, which means that each model in the suite should provide a routine that implements the (normalised) density $\rho(\bfx)$, and a routine that generates random positions according to this density. For example, the interface of a concrete {\tt{ConcreteGeometry}} class that contains only a single parameter $p$ would look as follows

{
\small
\begin{verbatim}
class ConcreteGeometry : public Geometry {
  public:
    ConcreteGeometry(double p) {_p=p}
    double density(Position x) const;
    Position generatePosition() const;
  private:
    double _p;
}
\end{verbatim}
}

\noindent 
For the design of this suite of input models, a naive option would be to provide a set of parameterised models that contain free parameters to control every possible option. A typical standard component of such a suite would be a very generalised Plummer model, with free parameters setting the location of the centre, the orientation with respect to the coordinate system, the scales, the flattening describing potential triaxiality, possible degrees of clumpiness, etc. This approach has a number of strong disadvantages. Clearly, it would quickly lead to an explosion of different options that is hard to overview. It would inevitably lead to substantial code duplication (nearly all the code would need to copied if we would consider a Hernquist profile instead of a Plummer profile) and code that is virtually impossible to maintain (the code for all models would need to be updated if a new feature is added or altered). Also concerning efficiency, such a design would not be optimal. Indeed, a random position generating routine for such a generalised Plummer model is hard to construct and is certainly much less efficient than the simple routine that is possible for a plain spherical Plummer model (using the inversion method). 

To overcome these problems, we adopted a completely different approach that is much simpler but still provides the flexibility and functionality needed to set up complex models. This is achieved using a combination of simple base models on the one hand, and so-called decorator geometries on the other hand. Decorator geometries are not real models on their own, but they apply modifications upon other models in interesting ways, following the {\em{Decorator}} design pattern. In object-oriented programming, a decorator attaches additional responsibilities to an object dynamically and provides a flexible and powerful alternative to subclassing for extending functionality \citep{gamma1994design}. 

In our present context, a decorator geometry is a special kind of geometrical model (i.e., it is also a C++ class in the general {\tt{Geometry}} suite) that takes one or more other components and adds a layer of complexity upon them. Simple examples of decorators that can easily be implemented are the relocation of the centre of a given component, or the rotation of a given model with respect to the coordinate system. More complex decorators deform a spherical model to a triaxial one, or add a spiral perturbation to an axially symmetric model. The advantage of the decorator approach is clear: each decorator needs to be implemented only once, and can then be applied to any possible model. 

In the following two sections we describe the building blocks in the SKIRT {\tt{Geometry}} suite, and a number of decorator geometries that can alter these building blocks to more complex structures.

\section{The SKIRT {\tt{Geometry}} building blocks}
\label{BuildingBlocks.sec}

\noindent
The SKIRT {\tt{Geometry}} suite contains a limited number of elementary input models, which can be used either as elementary toy models, or as building blocks for more complex geometries. For each of them, the density can be expressed as a simple analytical function, and the generation of random positions reduces to three independent univariate generation problems. 

Apart from these analytical components, SKIRT also offers the possibility to set up components in which the geometry of sources and/or sinks is defined by means of particles or on a grid. In particular, SKIRT can import a snapshot from a  (magneto)hydrodynamical simulation. One obvious goal would be to post-process a hydrodynamical simulation in order to calculate the observable multi-wavelength properties of the simulated objects \citep[e.g.,][]{2010MNRAS.403...17J, 2014MNRAS.445.3878S}. In this case, it would be sufficient to just read in the geometry of the snapshot as it is, and start the radiative transfer simulation. More generally, however, it would be useful if these numerical components would be at a similar level as the analytical components. This would open up the possibility to decorate them and combine them with other analytical and/or numerical components to more complex models. 

\subsection{Analytical components}
\vspace{0.5em}

\noindent 
A first group consists of spherically symmetric models. In spherical symmetry, the generation of a random position from the 3D density $\rho(\bfx) = \rho(r)$ simplifies to the generation of a random azimuth from a uniform distribution, a random polar angle from a sine distribution, and a random radius from the univariate density $f(r) = 4\pi r^2\rho(r)$. The SKIRT suite includes the most popular models used to represent shells, star clusters, early-type galaxies, or galaxy bulges, such as a power-law shell model \citep{1997MNRAS.291..121I}, the Plummer model \citep{1911MNRAS..71..460P}, the $\gamma$-model \citep{1993MNRAS.265..250D, 1994AJ....107..634T}, the S\'ersic model \citep{1963BAAA....6...41S, 1999A&A...352..447C, 2011A&A...525A.136B}, and the Einasto model \citep{1965TrAlm...5...87E, 2012A&A...540A..70R}. Within this family of models, many other famous profiles are contained, including the Hernquist model \citep{1990ApJ...356..359H}, the Jaffe model \citep{1983MNRAS.202..995J} and the $R^{1/4}$-model \citep{1948AnAp...11..247D}. 

A second group of elementary input models consists of axisymmetric models in which the density is a separable function. The standard example with a density distribution separable in cylindrical coordinates is the double-exponential disc that is the de facto standard to represent disk galaxies in radiative transfer studies \citep{1997A&A...325..135X, 2003MNRAS.343.1081B, 2007A&A...471..765B}. How random positions can be drawn from this distribution is discussed in Appendix~A of \citet{2003MNRAS.343.1081B}. An example of an axisymmetric model where the density is separable in spherical coordinates is the torus model that is often used to represent the dusty tori around stars or active galactic nuclei \citep{1991ApJ...368..545C, 1994MNRAS.268..235G, 2012MNRAS.420.2756S}.

\subsection{Components based on smoothed particles}
\label{SPH.sec}
\vspace{0.5em}

\noindent
The first group of numerical components in SKIRT are defined as a set of smoothed particles. This is mainly useful when we want to use the output of a smoothed particle hydrodynamics \citep{1977AJ.....82.1013L, 1992ARA&A..30..543M, 2010ARA&A..48..391S} simulation. In spite of claims that the technique suffers from fundamental problems \citep{2007MNRAS.380..963A, 2012MNRAS.423.2558B}, it is still the most popular hydrodynamics technique, especially for cosmological simulations of galaxy formation \citep[e.g.,][]{2011ApJ...742...76G, 2015MNRAS.446.1939F, 2015MNRAS.446..521S}. The output of an SPH snapshot consists of a set of ``particles'' (or rather anchor points in a co-moving grid), each characterised by a large suite of physical quantities. 

As far as SKIRT is concerned, most of these physical quantities are irrelevant. An {\tt{SPHGeometry}} component in SKIRT is essentially defined by a list of $N$ smoothed particles and the assumed smoothing kernel $W(r,h)$, with each smoothed particle characterised by a position $\bfx_j$, a fractional mass $m_j$ and a smoothing length $h_j$. The total density at an arbitrary position $\bfx$ is then given by
\begin{equation}
\rho(\bfx) = \sum_{j=1}^N m_j W(|\bfx-\bfx_j|,h_j)
\label{rho-SPH}
\end{equation}
In practice, the kernels used in SPH simulations almost always have a finite support \citep[e.g.,][]{1985A&A...149..135M, Desbrun1996, Muller2003}, and in this case only a relatively small number of terms in the sum have a non-zero contribution.

The {\tt{SPHGeometry}} class in SKIRT employs smoothing kernel implementations optimised for each specific task. The geometry's density at a given location is calculated according to equation (\ref{rho-SPH}) using a finite-support cubic spline kernel, so that the operation can be limited to particles that potentially overlap the location of interest. To facilitate this process, the setup phase of the simulation places a rough grid over the spatial domain and constructs a list of overlapping particles for each grid cell. As described in section 3.3 of \citet{2015A&C.....9...20C}, a further optimisation is provided to calculate the mass within a given box (a cuboid lined up with the coordinate axes), as an alternative to sampling the density in various locations across the volume of the box. In this case, the calculation uses the analytical properties of a scaled and clipped Gaussian kernel, designed to approximate the cubic spline kernel, to directly determine the mass in the box. This optimisation accelerates the density calculation for typical cartesian grids by an order of magnitude.

On the other hand, the generation of random positions sampled from the geometry's density distribution is rather straightforward, thanks to the composition method. The first step is the choice of a random smoothed particle, based on a discrete distribution where each particle is weighted by its relative mass contribution. The second step is generating a random position according to the distribution of the chosen particle. The current implementation samples a random number from a Gaussian smoothing kernel with infinite support using the inversion method.

In the future we may provide a suite of smoothing kernels, allowing the user to select the kernel that best fits the SPH snapshot being imported. The methods used for calculating the density and generating a random radius would obviously depend on the actual shape of the kernel.

In addition to what is described above, SKIRT has facilities to associate a spectral energy distribution with each SPH star particle based on properties such as age and metallicity, or to associate a dust mass with each SPH gas particle based on properties such as temperature and metallicity. In the former case, the geometry will weigh the particles by luminosity, for each wavelength separately, and in the latter case, the geometry will weigh the particles by dust mass. A detailed description of these features is beyond the scope of this paper.

\subsection{Components based on hierarchical grids}
\vspace{0.5em}

\noindent
Apart from smoothed particle hydrodynamics, the main other technique that is used to perform hydrodynamics simulations is Eulerian mesh-based hydrodynamics \citep{1992ApJS...80..753S, 2000ApJS..131..273F}. A fundamental ingredient of this technique is the use of grids with adaptive mesh refinement. Eulerian AMR simulations are used in virtually all fields of astrophysics, and Monte Carlo codes have been adapted to work directly on these hierarchical grids \citep{2001A&A...379..336K, 2006A&A...456....1N, 2013A&A...554A..10S, 2014A&A...561A..77S}.

The {\tt{HierarchicalGridGeometry}} in SKIRT reads in snapshots defined by density fields discretised on hierarchical cartesian grids and converts them to the format of the other components in the SKIRT {\tt{Geometry}} suite. Calculating the density at an arbitrary position comes down to identifying the cell that contains this position and returning the density associated with that cell. As hierarchical grids have the structure of a tree, isolating the correct cell is a straightforward and computationally cheap operation. Generating random positions from such a component is similar to the case of the smoothed particles, and is based on the composition method. We first generate a random cell from a discrete distribution where each cell is weighted by its relative mass. Secondly, we determine a random position within the chosen cell; as the cells in a hierarchical cartesian grids are cuboids, this is a trivial task.

\subsection{Components based on Voronoi grids}
\vspace{0.5em}

\noindent
Recently, a new Lagrangian technique that solves the hydrodynamics equations on a moving, unstructured Voronoi grid is gaining popularity. It is claimed to avoid some of the difficulties of smoothed particle hydrodynamics on the one hand and Eulerian grid-based hydrodynamics on the other hand. This technique has been used for many years in the computational fluid dynamics community \citep{1997AnRFM..29..473M}, and a number of novel codes based on this principle have recently been developed in the astrophysics community \citep{2010MNRAS.401..791S, 2011ApJS..197...15D}. Moving mesh hydrodynamics is mainly applied to simulations of galaxy formation and evolution \citep[e.g.,][]{2014MNRAS.437.1750M, 2014MNRAS.444.1518V}.

SKIRT contains a {\tt{VoronoiGridGeometry}} class that converts a snapshot from a Voronoi hydrodynamical simulation (or any other density field defined on a Voronoi grid) to a regular SKIRT {\tt{Geometry}} component. Due to the nature of a Voronoi grid, the only necessary input is the list of all the generating sites and the associated densities; it is hence not necessary to store all the vertices and sides of each of the cells. Based on the generating sites, SKIRT constructs the corresponding unique Voronoi grid using the public Voro++ library \citep{Rycroft2009}.

The density routine essentially works in the same way as for the case of the hierarchical grids: it comes down to identifying the cell that contains the given position and returning the density associated to this cell. In the case of a Voronoi grid, however, the cell identification is not as simple as in a hierarchical tree structure of cartesian grid cells. Due to the nature of Voronoi grids, this is essentially a nearest neighbour search. Rather than looping over all possible sites, SKIRT implements an approach using cuboidal blocks, as explained in detail in \citet{2013A&A...560A..35C}. This task could be optimised even further using more advanced techniques based on space partitioning structures such as $k$-d trees or R-trees \citep{Friedman:1977:AFB:355744.355745, Guttman:1984:RDI:971697.602266, Liaw:2010:FEK:1746725.1746801}.

Also the generation of random positions works essentially in the same way as for hierarchical grids. The first step is identical: we generate a random cell from a discrete distribution where each cell is weighted by its relative mass contribution. The second step, generating a random position from within the chosen cell, is significantly more complex than in the case of a cuboidal cell. To the best of our knowledge, there are no dedicated techniques to generate a random point from a Voronoi cell. There are two possible options. 

The first option is to partition the cell into a set of tetrahedra, subsequently select a random tetrahedron from a discrete distribution where every tetrahedron is weighted by its relative volume, and finally generate a random position from the selected tetrahedron. Specific algorithms are available for both the tetrahedrisation of convex polyhedra \citep{Edelsbrunner1990335, Max:2002:CSC:607954.607957} and the generation of random positions from a tetrahedron \citep{Rocchini:2000:GRP:378572.378603}. 

The second option, which is more simple and which we have adopted in SKIRT, is to use the rejection technique. As the reference distribution we use a uniform density in a cuboidal volume, defined as the 3D bounding box of the cell. As Voronoi cells are convex polyhedra, this bounding box is directly obtained when the vortices of the cell are known (these are calculated using the Voronoi grid setup). Extensive testing has shown that, depending on the distribution of the generating Voronoi sites, the average ratio of the volume of the bounding box of a Voronoi cell over the actual cell volume is about 3 to 4. This ratio immediately represents the average rejection rate for the random position generation.

\section{The SKIRT {\tt{Geometry}} decorators}
\label{Decorators.sec}

\begin{table}
  \caption{An overview of the geometry decorators currently implemented in SKIRT, referencing the section in which they are presented. Geometry decorators can be applied to basic geometry building blocks, and can be chained and combined to create complex structures.}
  \label{decorators.tab}
  \centering
  \footnotesize
  \begin{tabular}{p{0.03\linewidth}p{0.15\linewidth}p{0.66\linewidth}}
    \\ \hline
      \ref{Offsets.sec} & {\tt{Offset}} & applies an arbitrary offset to any geometry \\
      \ref{Rotation.sec} & {\tt{Rotate}} & applies an arbitrary rotation to any geometry \\
      \ref{Cavities.sec} & {\tt{SpheCavity}} & carves out a spherical cavity from any geometry\\
                                     & {\tt{Crop}} & crops any geometry to a given box; a variation on the {\tt{Cavity}} decorator \\
      \ref{Composition.sec} & {\tt{Combine}} & combines two geometries into a single geometry \\
      \ref{Triaxial.sec} & {\tt{Triaxial}} & transforms any spherical geometry to a triaxial form \\
                                    & {\tt{Spheroidal}} & transforms any spherical geometry to a spheroidal form; a special case of the {\tt{Triaxial}} decorator  \\
      \ref{SpiralArmStructure.sec} & {\tt{Spiral}} & applies spiral arm structure to any axisymmetric geometry \\
      \ref{ClumpyModels.sec} & {\tt{Clumpy}} & replaces a portion of the mass in any geometry by randomly placed clumps \\
      \ref{GenericRandom.sec} & {\tt{Foam}} & replaces the model-specific random position generator by a generic routine based on the Foam library \\
    \hline
  \end{tabular} 
\end{table}

\noindent
In this section we describe a number of decorator geometries that can be applied on the building blocks described in the previous section in order to convert them to more complex structures; see Table~\ref{decorators.tab} for an overview.  The implementation of the density of a decorator geometry is usually not a major problem; the main challenge is to implement the routine that generates random positions from a decorator geometry, so this is what we focus on.

\subsection{Offsets}
\label{Offsets.sec}
\vspace{0.5em}

\noindent
The {\tt{OffsetGeometryDecorator}} decorator in SKIRT applies an arbitrary offset $\bfa$ to any density distribution. If the original density is $\rho_\txs(\bfy)$, the new density is simply $\rho(\bfx) = \rho_\txs(\bfx-\bfa)$. Generating random positions is equally simple: we generate a random $\bfy$ from the original density distribution and return $\bfx = \bfy+\bfa$. The C++ implementation of the class in SKIRT looks like:

{
\small
\begin{verbatim}
class OffsetGeometryDecorator : public Geometry {
  public:
    OffsetGeometryDecorator(Geometry* g, Position a) 
      {_g = g; _a = a;}
    double density(Position x) const 
      {return _g->density(x-a);}
    Position generatePosition() const 
      {return _g->generatePosition()+a;}
  private:
    Geometry* _g;
    Position _a;
}
\end{verbatim}
}

\subsection{Rotation}
\label{Rotation.sec}
\vspace{0.5em}

\noindent
Similarly, the {\tt{RotateGeometryDecorator}} decorator rotates any density distribution. If the rotation is characterised by the orthonormal matrix ${\bf{H}}$, the new density is $\rho(\bfx) = \rho_\txs({\bf{H}}\,\bfx)$. To generate a random position from this new density, we generate a random $\bfy$ from the original density and rotate it over the inverse rotation matrix, i.e. $\bfx = {\bf{H}}^{\text{T}}\bfy$.

\subsection{Cavities}
\label{Cavities.sec}
\vspace{0.5em}

\noindent
A third simple decorator, the {\tt{CavityGeometryDecorator}}, car\-ves out a cavity from another density $\rho_\txs(\bfy)$. In formula form, we have
\begin{equation}
\rho(\bfx) = 
\begin{cases}
\;0 & \quad\text{$\bfx$ in cavity} \\
\;\dfrac{\rho_\txs(\bfx)}{1-\chi} & \quad\text{else}
\end{cases}
\end{equation} 
with $\chi$ the fraction of the mass of the original density located in the cavity. This decorator is useful to represent density distributions of dust close to a star or active galactic nucleus, where the dust has been cleared due to sublimation. To generate random positions from this new density distribution, we just generate a random position from the original density distribution and reject is when it is located in the cavity. This is, in fact, an almost trivial application of the rejection technique, where the original density assumes the role of the reference function, and the rejection constant is $c = 1/(1-\chi)$.

\subsection{Composition}
\label{Composition.sec}
\vspace{0.5em}

\noindent
The {\tt{CombineGeometryDecorator}} combines two density distributions into a single distribution according to
\begin{equation}
\rho(\bfx) = \frac{w_1\,\rho_1(\bfx)+w_2\,\rho_2(\bfx)}{w_1+w_2} 
\end{equation} 
with $\rho_1,\rho_2$ the original distributions and $w_1,w_2$ their respective weights in the composite distribution. Generating random positions for this new density distribution is a trivial application of the composition method.

\subsection{Triaxial geometries}
\label{Triaxial.sec}
\vspace{0.5em}

\noindent
As a first more complex decorator, we consider the case of a triaxial decorator geometry, which converts a spherically symmetric density distribution into one that is stratified on concentric and confocal ellipsoids. More concretely, assume that we have a spherically symmetric density distribution $\rho_\txs(\bfy) = \rho_\txs(y)$, we then consider its triaxial counterpart
\begin{equation}
\rho(\bfx)
=
\frac{1}{pq}\,\rho_\txs\left(\sqrt{x_1^2 + \frac{x_2^2}{p^2} + \frac{x_3^2}{q^2}}\right)
\label{tri:density}
\end{equation}
Such triaxial models are discussed and used extensively to describe the stellar distribution of elliptical galaxies and galactic nuclei and the shape of galactic haloes \citep[e.g.,][]{1977ApJ...213..368S, 1996ApJ...460..136M, 2002MNRAS.333..510T, 2008MNRAS.385..647V}. Oblate and prolate spheroidal distributions are just a special case of these triaxial models in which $p=1$.

It is clear that the density (\ref{tri:density}) cannot be written as a product of independent univariate density distributions. We can use the conditional distribution method. In order to develop a general recipe for generating random positions, we start by rewriting the probability distribution according to~(\ref{tri:density}) in spherical coordinates, 
\begin{equation}
f(r,\theta,\phi)
=
\frac{r^2\sin\theta}{pq}\,\rho_\txs\left(
r\sqrt{\frac{\sin^2\!\theta\,(\sin^2\!\phi+p^2\cos^2\!\phi)}{p^2}+\frac{\cos^2\!\theta}{q^2}}
\right)
\end{equation} 
According to formula~(\ref{cdm}), we now rewrite this expression as
\begin{equation}
f(r,\theta,\phi) = f(\phi)\,f(\theta\,|\,\phi)\,f(r\,|\,\theta,\phi)
\end{equation}
After some calculation, one finds for the marginal distribution for $\phi$ the simple expression
\begin{equation}
f(\phi)
=
\frac{1}{2\pi}\,\frac{p}{\sin^2\!\phi+p^2\cos^2\!\phi}
\label{tri:f(phi)}
\end{equation}
Note that this marginal distribution is independent of the specific shape of the density profile and only depends on the flattening parameter $p$ (and for $p=1$, it reduces to a simple uniform distribution, as expected for a spheroidal distribution). Generating a random azimuth from this density can be done using the standard inversion method; the corresponding cumulative distribution is
\begin{equation}
F(\phi)
=
\frac{1}{2\pi}\arctan\left(\frac{\tan\phi}{p}\right)
\end{equation}
and is readily inverted. Once this random $\phi$ has been determined, we can generate a random polar angle $\theta$ from the conditional distribution 
\begin{equation}
f(\theta \,|\, \phi)
=
\frac{\sin\theta}{2s}\left(\frac{\sin^2\!\theta}{s^2}+\cos^2\!\theta\right)^{-3/2}
\label{tri:f(theta)}
\end{equation}
where we have set $s^2 = q^2(\sin^2\!\phi+p^2\cos^2\!\phi)/p^2$. Also this distribution is independent of the specific shape of the density profile and only depends on the flattening parameters $p$ and $q$ (and the already randomly determined azimuth). Again, generating a random polar angle from this distribution can be done using the standard inversion method: the corresponding cumulative distribution is
\begin{equation}
F(\theta\,|\,\phi)
=
\frac12\left(1-\frac{1}{\sqrt{1+s^2\tan^2\!\theta}}\right)
\end{equation}
and is again readily inverted. Finally, we consider the conditional distribution for $r$, 
\begin{equation}
f(r\,|\,\theta,\phi)
=
\frac{4\pi r^2}{q^3}
\left(\frac{\sin^2\!\theta}{s^2}+\cos^2\!\theta\right)^{3/2}\,
\rho_\txs\left(\frac{r}{q}\sqrt{\frac{\sin^2\!\theta}{s^2}+\cos^2\!\theta}\right)
\end{equation}
If we apply a linear transformation from $r$ to 
\begin{equation}
y
=
\frac{r}{q}\sqrt{\frac{\sin^2\!\theta}{s^2}+\cos^2\!\theta}
\label{mr}
\end{equation}
the corresponding density for $m$ is simply
\begin{equation}
f(y\,|\,\theta,\phi)
=
4\pi y^2\rho_\txs(y)
\end{equation}
which is nothing but the distribution for a random radius from the original density $\rho_\txs(\bfy)$. The last step in the generation of a random position is to generate a random position $\bfy$ from the original spherical density distribution $\rho_\txs(\bfy)$, and transform the radius $y=|\bfy|$ of this position to a new radius $r$ using the linear transformation~(\ref{mr}), in which we use the previously determined values for $\theta$ and $\phi$.

While this method based on conditional probabilities has a certain beauty, there is, actually, a simpler and more efficient method that is based on the inversion method. Indeed, consider the simple transformation
\begin{equation}
(y_1,y_2,y_3) = (x_1,x_2/p,x_3/q)
\end{equation}
or in matrix format
\begin{equation}
\bfy = {\bf{H}}\,\bfx
\qquad
{\bf{H}} = \begin{pmatrix} 1 & 0 & 0 \\ 0 & \frac1p & 0 \\ 0 & 0 & \frac1q \end{pmatrix}
\end{equation}
According to the law of linear transformations (\ref{transfmethodlin}), if $\bfy$ is distributed according to the spherical density $\rho_\txs(\bfy)$, $\bfx$ will be distributed exactly according to the triaxial density~(\ref{tri:density}). An easy way to generate random positions from a triaxial counterpart of a spherical symmetric density is thus simply generating a random position $\bfy$ from the spherical density distribution, from which the desired position can easily be calculated as $\bfx = {\bf{H}}^{\text{T}}\bfy$. 

\subsection{Spiral arm structure}
\label{SpiralArmStructure.sec}
\vspace{0.5em}

\noindent
As a second complex decorator, we consider a logarithmic spiral arm perturbation that can be applied to any axisymmetric density distribution, 
\begin{subequations}
\label{spiral}
\begin{equation}
\rho(\bfx) = \rho_\txs(R,z) \left[(1-w) + w\,\xi(R,\phi)\right]
\end{equation}
where $(R,\phi,z)$ are the usual cylindrical coordinates, and the function $\xi(R,\phi)$ is defined as
\begin{equation}
\xi(R,\phi) = C_N \sin^{2N} \left[\frac{n}{2}\left( \frac{\ln (R/R_0)}{\tan p}-\phi\right) + \frac{\pi}{4} \right]
\label{perturbation}
\end{equation}
\end{subequations}
The factor $C_N$ is a normalisation constant that guarantees that the density is normalised, and is given by 
\begin{equation}
C_N = \frac{1}{2\pi} \int_0^{2\pi} \sin^{2N}\!\phi\,\txd\phi = \frac{\sqrt{\pi}\,\Gamma(N+1)}{\Gamma(N+\frac12)}
\end{equation}
This general spiral arm perturbation is a generalisation of the models used by \citet{2000A&A...353..117M} and \citet{2012ApJ...746...70S}, and allows for an arbitrary number of arms $n$, pitch angle $p$, spiral perturbation weight $w$, and arm-interarm size ratio (controlled by the parameter $N$). For $N=1$ the density reduces to a more simple perturbation with equal arm-interarm size ratio,
\begin{equation}
\rho(\bfx) = \rho_\txs(R,z) \left\{1 + w\sin \left[n\left( \frac{\ln (R/R_0)}{\tan p}-\phi \right) \right] \right\}
\end{equation}
For the general 3D density distribution (\ref{spiral}), we can again use the conditional distribution method, and write
\begin{equation}
f(R,\phi,z) = f(R,z)\,f(\phi\,|\,R,z)
\end{equation}
Since the spiral perturbation is such that it averages out along every single circle around the $z$-axis, i.e.,
\begin{equation}
\int_0^{2\pi} \xi(R,\phi)\,\txd\phi =1
\end{equation}
we trivially find 
\begin{equation}
f(R,z) = 2\pi R\,\rho_\txs(R,z)
\end{equation}
This is exactly the 2D probability density distribution that corresponds to the density of the axisymmetric density distribution. We hence simply generate a random position $\bfx$ from our original density distribution and save the radius $R$ and the height $z$. In order to generate a random azimuth $\phi$, we consider the conditional distribution $f(\phi\,|\,R,z)$, for which we find directly
\begin{equation}
f(\phi\,|\,R,z)
=
\frac{(1-w) + w\,\xi(R,\phi)}{2\pi}
\label{spiral:PDF}
\end{equation}
This univariate density is independent of the shape of the original axisymmetric density profile; it only depends on the parameters of the spiral perturbation and randomly determined radius $R$. For the general expression~(\ref{perturbation}), the standard inversion technique is not easily applicable; the cumulative distribution corresponding to the density~(\ref{spiral:PDF}) can be calculated analytically, but the resulting expression cannot be inverted analytically. One could resort to a numerical inversion, but a better approach is to use the rejection technique. We can use a simple uniform density as the reference distribution, $f_{\text{ref}}(\phi) = 1/2\pi$. The rejection constant $c$ should ideally be chosen as the smallest value for which $f(\phi\,|\,R,z) \leqslant cf_{\text{ref}}(\phi)$ for all values of $\phi$. Since the maximum value of the perturbation function $\xi(R,\phi)$ is $C_N$, we find
\begin{equation}
c = 1 + (C_N-1)\,w
\end{equation}
Actually, as an alternative to the conditional probability technique, we could also simply have used the rejection technique from the beginning. The 3D density $\rho(\bfx)$ satisfies the condition $\rho(\bfx) \leqslant c \rho_\txs(\bfx)$, so if we simply use the original density $\rho_\txs(\bfx)$ as our reference distribution, we have essentially the same algorithm with the same rejection constant. The difference between this 3D rejection technique and the previous version, where we used the rejection technique only for the conditional distribution for the azimuth, is a matter of efficiency. In the former version only one random position needs to be generated from the original density, and it the rejection is only to determine the azimuth. In the latter version, an entirely new position is generated for every rejection, which is less efficient. 

\subsection{Clumpy models}
\label{ClumpyModels.sec}
\vspace{0.5em}

\noindent
As a last example, we consider the {\tt{ClumpyGeometryDecorator}} class that turns any density distribution $\rho_\txs(\bfy)$ into a clumpy analogue, i.e., a density distribution in which a fraction of the mass is distributed ``smoothly'' as in the original density distribution, and the remaining fraction is locked up into compact clumps. The density of this distribution can be written as
\begin{equation}
\rho(\bfx) = (1-f_{\text{cl}})\,\rho_\txs(\bfx) + \frac{f_{\text{cl}}}{N}\sum_{i=1}^N W(|\bfx-\bfx_i|,h)
\label{clumpy}
\end{equation}
where $f_{\text{cl}}$ is the fraction of the mass in clumps, $N$ is the number of clumps, and the positions $\bfx_i$ are the central positions of the $N$ clumps, determined randomly according to the original density function $\rho_\txs$. The function $W(r,h)$ is the smoothing function that sets the distribution of matter within every single clump, with $h$ a characteristic length scale. This can in principle be any spherical 3D distribution; in practice we use either a uniform density sphere or one of the compact support kernels that are used in smoothed particle hydrodynamics simulations.

It is straightforward to generate random positions from this distribution. It is a direct application of the composition method, and the case is very similar to smoothed particle building blocks discussed in Section~{\ref{SPH.sec}}. 

\section{Discussion}
\label{Discussion.sec}

\subsection{Advantages of the decorator-style design}
\vspace{0.5em}

\noindent
As indicated in Section~{\ref{GeometrySetup.sec}}, a major advantage of the decorator-style approach is that each decorator needs to be implemented only once, and can then be applied to different models. This avoids the need to implement complex geometries with many different possible features and heavy code duplication. That alone is already an argument strong enough to justify its use.

\begin{figure*}[t!]
\centering
\includegraphics[width=0.75\columnwidth]{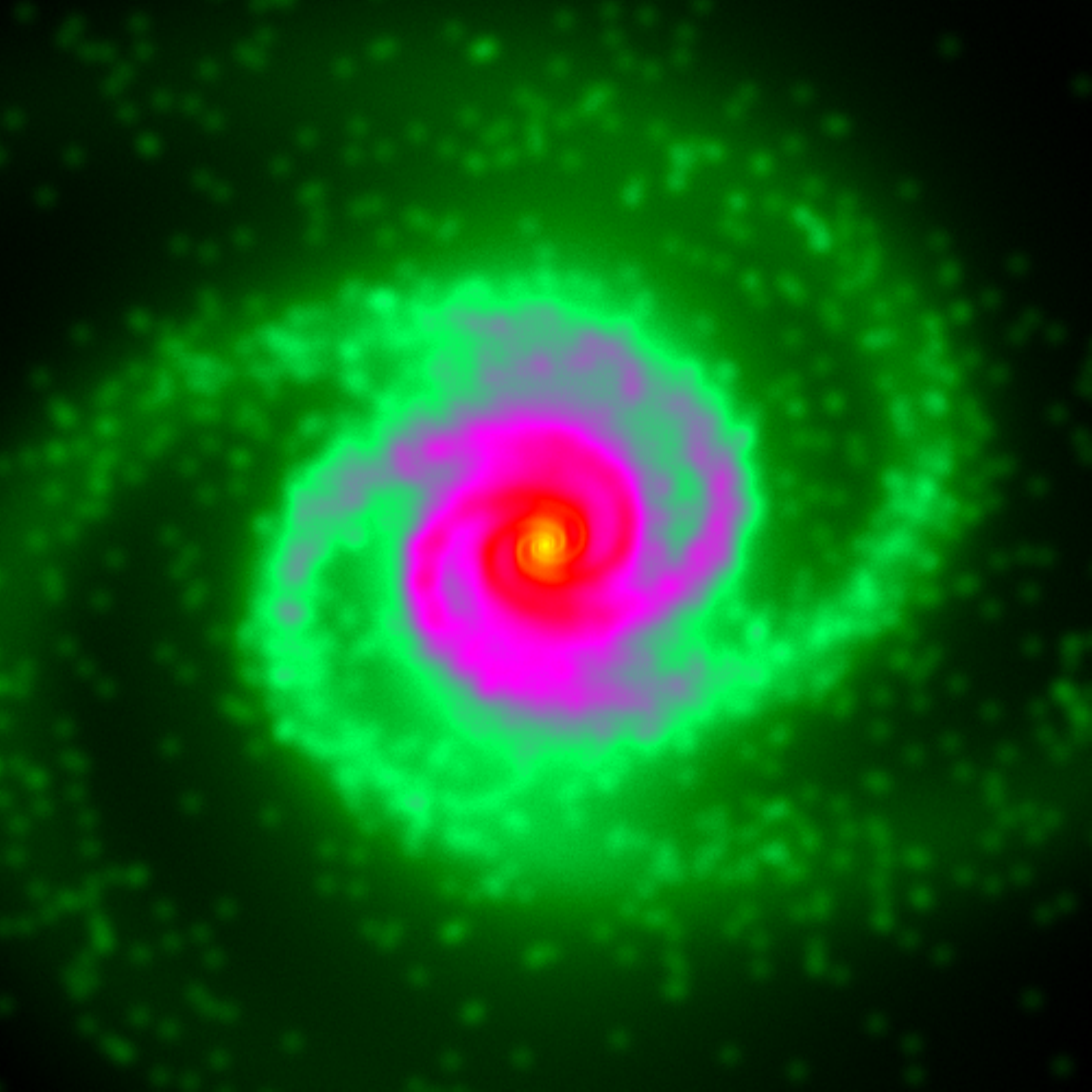}
\includegraphics[width=0.75\columnwidth]{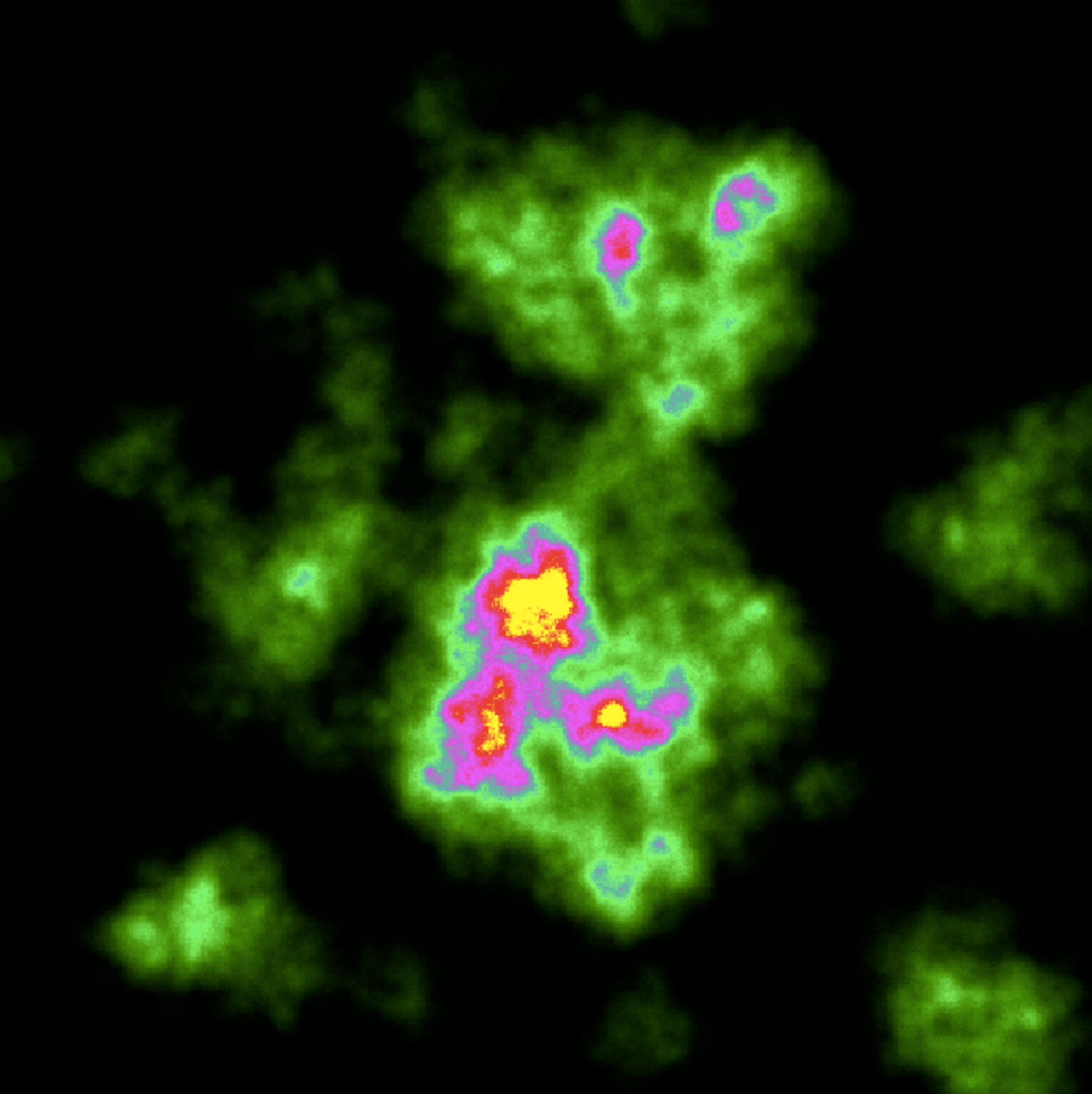}
\caption{Two examples of the chained application of decorators. The left panel is a simulated image of a two-component model consisting of a triaxial bulge, and a flattened disc model on which a three-armed logarithmic spiral perturbation and a clumping decorator were applied. For both the bulge and the disc component, the starting model was a simple spherical $\gamma$ model. The model to the right is a triaxial Plummer model, with four recursive applications of the {\tt{ClumpyGeometryDecorator}} decorator. The result is a model with a fractal structure. In both models $10^9$ photon packages were released to create the images, and the noise in the images is negligible (i.e., all the features in the image are real).}
\label{ChainedDecorators.fig}
\end{figure*}

Another strong advantage is the flexibility offered by this approach. The decorators discussed in Section~{\ref{Decorators.sec}} transform an input model into a more complex model. Such a decorated model is often still relatively simple: it typically adds one layer of complexity onto the model that is being decorated (think of a rotated exponential disc or a triaxial S\'ersic model). If we want to generate toy models that can be used as idealised representations of, for example, a spiral galaxy, various layers of complexity have to be added. The decorator-style approach is ideal for this purpose: decorators can be applied also to models that have already been decorated. In other words, decorators can easily be chained or nested. 

The left-hand panel of Figure~{\ref{ChainedDecorators.fig}} shows a projected image of a relatively complex toy model, constructed using the {\tt{Geometry}} suite in SKIRT. The model consists of a bulge component and a clumpy spiral disc. Each of these two components have a very simple spherical $\gamma$-model as their starting point, on which a chain of decorators was applied. The former was first decorated into a triaxial model, the latter was turned into an axisymmetric disc by applying a very strong flattening, then a spiral perturbation was added, and a fraction of its mass was turned into clumps. 

In decoration chains the successive decorators do not necessarily need to be different. It is also possible to chain the same decorator several times, i.e.\ to apply them recursively. A powerful example of this nesting is a repeated application of the {\tt{ClumpyGeometryDecorator}}. The right-hand panel of Figure~{\ref{ChainedDecorators.fig}} shows an application of this principle on a simple triaxial Plummer model. In this case, we have applied four nested applications of the decorator, in which the number of clumps increases in each level, whereas the smoothing length decreases. This algorithm results in a fractal density distribution that is self-similar over an order of magnitude in scale \citep{1997ApJ...477..196E, 2002ApJ...574..812M, 2006ApJ...636..362I}. 

The possibility to chain decorators, including a repeated application of the same decorator, facilitates the construction of very complex models out of simple building blocks. Analytical components can easily be combined with numerical components based on smoothed particles or grids, and decorators can be applied regardless of the underlying component type. It is, for example, possible to add a smooth triaxial bulge to an irregular disc structure defined using SPH particles, or to carve out a cavity from a complex hydrodynamics system and locate a small nuclear structure there to simulate the effects of an AGN on the large-scale structure of a galaxy.

\subsection{The use of a generic random position generator}
\label{GenericRandom.sec}
\vspace{0.5em}

\begin{figure}
\centering
\includegraphics[width=0.82\linewidth]{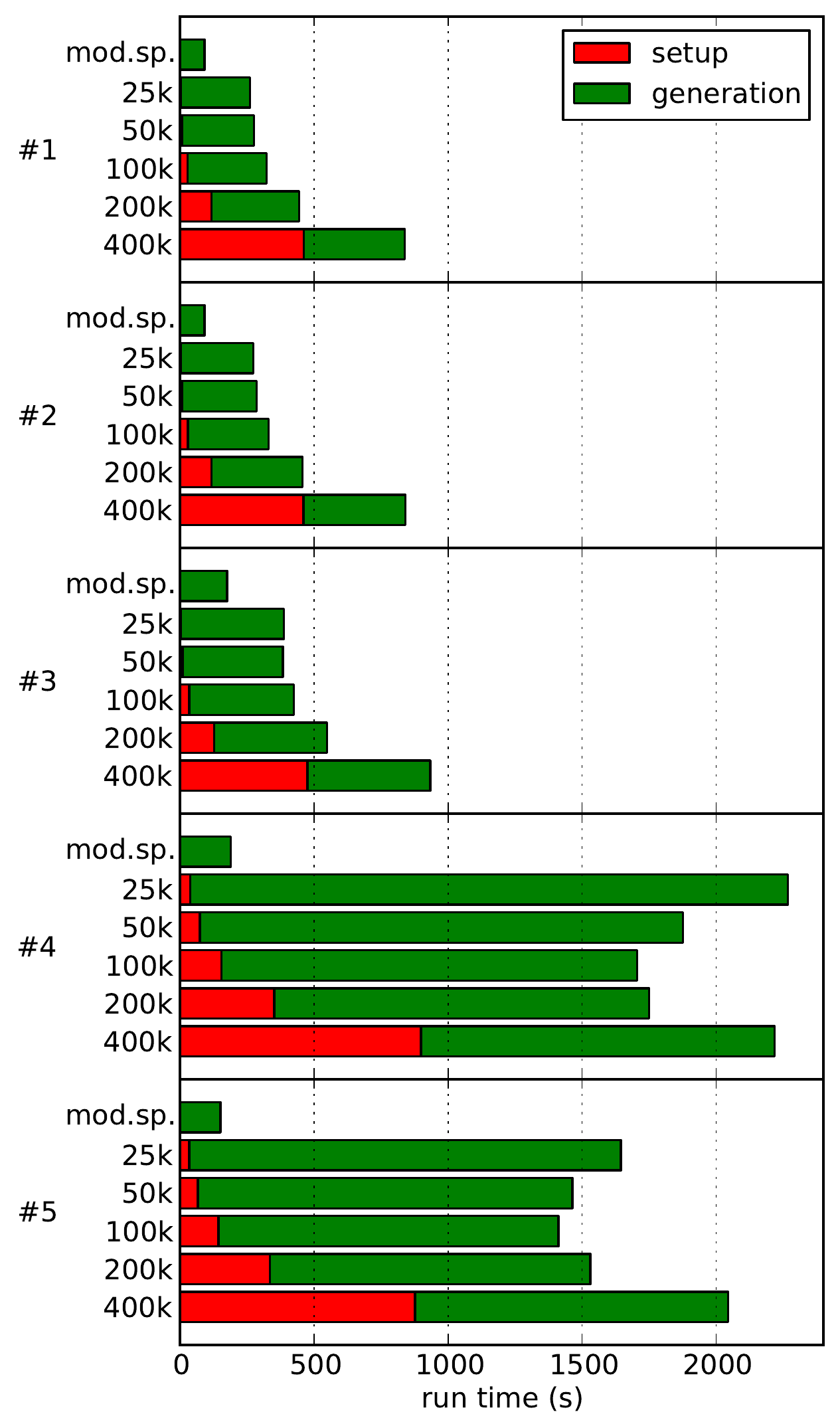}
\caption{Results of the timing tests for the test models \#1 through \#5, using the model-specific and the generic Foam-based random position generators with 25,000 through 400,000 cells. The horizontal bars show the simulation run-time in seconds, split into the fraction devoted to the setup of the grid (red) and the actual time spent on random position generation (green).}
\label{timings.fig}
\end{figure}

\noindent
One could argue that the setup that we have chosen is unnecessarily complex. An alternative approach would be a design where we only provide the density for each component in the suite, and where the generation of random positions is executed by a generic routine rather than a geometry-specific routine for each component. Such a suite could still use the decorator design pattern, but would be simpler to implement. The main challenge is the design of a generic routine that can generate random positions corresponding to an arbitrary 3D density function. 

This is in principle possible: there are a few multivariate black-box random number generation libraries available, including Foam \citep{Jadach2000, Jadach2003} and RanLip \citep{Beliakov2005b, Beliakov2005}, based on the so-called composition-rejection method, a combination of the composition and rejection methods. In a first phase, the exploration phase, the domain of a distribution is partitioned into different cells. In the generation phase, the rejection technique is used on each cell, where typically a constant density function is used as the reference distribution. The advantages of this approach are that, in principle, it can be applied for all distributions and in any dimension, and that only the density of each component is required. The main disadvantage is the complexity and the overhead, both in memory and run-time, that are linked to the exploration phase. 

In order to test whether or not our design choice of providing a customised random position generator for each concrete {\tt{Geometry}} class (and for each decorator, in particular) is justified, we have set up a comparison between our routines and a generic black-box routine. We have implemented a new decorator class, {\tt{FoamGeometryDecorator}}, that replaces the model-specific random position generator by a generic routine based on the Foam library. Foam is a self-adapting cellular code that uses iterative binary splitting, using either simplexes or hyper-rectangles, to subdivide the configuration space. 

We set up a simple suite of test models, for which we generate random positions using both the model-specific generator and the generic Foam generator. Our suite consists of a sequence of models with an increasing level of complexity. It essentially follows the different steps in building the model shown in the left panel of Figure~{\ref{ChainedDecorators.fig}}. The following models are considered
\begin{enumerate}
\item a spherical $\gamma=\tfrac12$ model, 
\item model \#1 flattened to an axisymmetric disc,
\item model \#2 with spiral perturbation,
\item model \#3 with a fraction turned into clumps,
\item model \#4 with a triaxial $\gamma=\tfrac12$ bulge added.
\end{enumerate} 
For each model, we generate $10^8$ random positions (we do not use any photon propagation or detection as would be used in Monte Carlo radiative transfer simulation in order to isolate the random position generation process). The timings were done using a single core and with averages over multiple test runs. One crucial aspect for the Foam-based generator is the choice of the Foam parameters, in particular of the number of cells in the grid. A larger number of cells implies a higher computational cost of the exploration phase on the one hand, but a better approximation of the density and smaller rejection rates. The ideal choice of this parameter is impossible to determine in a general way. We consider 5 different values for the number of cells in the Foam for every model, ranging from 25,000 to 400,000. 

The results are illustrated in Fig.~{\ref{timings.fig}}, where, for each model, we give the total run time and the contributions for the setup of the foam and the actual time spent on generating the random positions. 

A first major conclusion that can be drawn from these results is the efficiency of the customised random position generation routines. From a simple spherical model, to a complex model composed of two components that both have been decorated multiple times, the computational cost increases only by a factor two. Moving from a spherical to a triaxial model does not affect the efficiency of the generation of random positions at all (not a surprise, as it just implies two simple multiplications, as shown in Section~{\ref{Triaxial.sec}}). The biggest jump in timing occurs when spiral structure is added to the model (i.e., from model \#2 to \#3), because the random position generation is based on the rejection method, which has more overhead compared to the inversion and composition methods. The addition of clumping (from model \#3 to \#4) hardly implies an increase in computation time. Note that the addition of a bulge even decreases the computation time. In model \#5 half of the positions are generated according to the density of the relatively simple triaxial bulge, whereas in model \#4 they all are generated according to the more complex clumpy spiral structure.

A second major conclusion is that the generic Foam-based random position generator is significantly less efficient than the customised generator. For the most simple models (\#1 and \#2), the difference in speed is a factor 3--8, depending on the number of cells used. The increase in total run time for the Foam-based models with increasing number of cells is mainly due to the time necessary to set up the foam. The time necessary to generate the random positions also increases, but not as strongly. 

\begin{figure*}[t!]
\centering
\includegraphics[width=\textwidth]{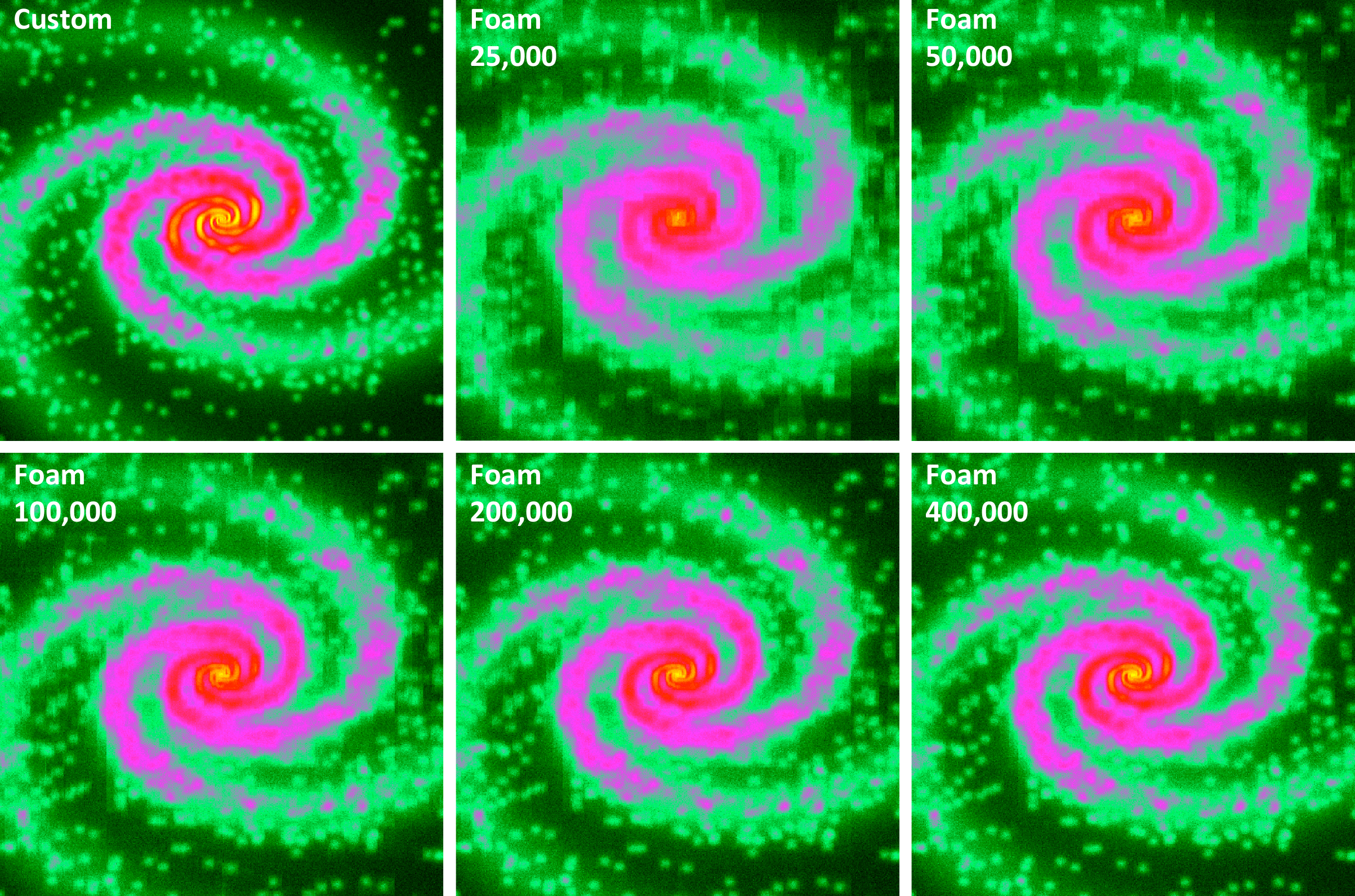}
\caption{A comparison of the Monte Carlo generated images for test model \#4, based on $10^8$ photon packages. For the top-left panel, we have used the customised random position generator based on the chain of decorators. For the other panels, we have used the generic Foam-based generator, with the number of cells in each panel indicated in the top-left corner.}
\label{GenericVersusFoam.fig}
\end{figure*}

For the more complex models, the efficiency of the Foam-based generator decreases even more. The biggest jump in timing now occurs when we add clumping to the model, because the calculation of the density for the clumpy models is computationally very demanding (each density evaluation requires a sum over a large number of clumps). For models \#4 and 5, the Foam-based generators are at least an order of magnitude slower than the customised routines. Interestingly, the run time of these simulations does not increase with the number of cells. For simulations with few cells, the setup phase is relatively quick, but the generation phase is very inefficient. The reason is that the cells are relatively large with a strong variation in density values, and hence the rejection rates are uncomfortably large. If the number of cells in the grid increases, the setup time naturally increases, but the actual position generation becomes more efficient because the average rejection rates are smaller.

Moreover, the inefficiency of the Foam-based generator is not the only problem, but also accuracy is an issue, in particular for complex models with multiple local maxima and strong gradients, such as models \#4 and \#5. If the number of cells is limited and the cells hence rather large, both the total mass and the local maximum within a cell are very hard to guess (they are determined by random sampling the cell). As a result, both the relative weights of the different cells and the rejection constants within the individual cells are poorly determined. This will lead to an artificial smoothing and wrong results. Figure~{\ref{GenericVersusFoam.fig}} shows a sequence of images corresponding to model \#4, created using the customised random position generator (top left) and using the Foam-based one with different values for the number of cells (remaining panels). The customised generator generates an image that reveals all the details that are expected, including the sharp density gradients and the contrast between arm and interarm regions. The only limitation in the image is the finite resolution due to the pixel size and the unavoidable Poisson noise. The Foam images on the other hand show a clear signature of degradation that gradually decreases when the number of cells grows. For the grids with 25,000 and 50,000 cells, and even for the one with 100,000 cells, the effects of the grid are clearly visible, and the individual clumps and the spiral structure are insufficiently resolved.  For the grid with 400,000 cells, the individual clumps are well resolved, but the sharpness of the spiral arms is still under-resolved, especially in the central regions.  

Since both the accuracy and the efficiency of the generic Foam-based random position generator cannot compete with the customised versions, we conclude that it is worth investing in the latter, and that our design choice is justified.

\section{Conclusions}
\label{Conclusions.sec}

\noindent
We have described the design of a suite of components that can be used to model the distribution of sources and sinks in the Monte Carlo radiative transfer code SKIRT. Our main conclusions are the following:
\begin{itemize}
\item 
The availability of a well-designed suite of input models, with enough variety and different degrees of complexity, in a publicly available Monte Carlo code has a strong added value. Such models can serve as toy models to test new physical ingredients, or as parameterised models for inverse radiative transfer fitting. They also provide a low-threshold introduction to new users, since models with differing degrees of complexity can be run without any coding at all.
\item
Each model is in principle completely determined by the 3D density distribution $\rho(\bfx)$. In order to be used in Monte Carlo radiative transfer, however, each model should contain a routine that generates random numbers according to this $\rho(\bfx)$. Finding the most suitable algorithm to implement this is the most challenging aspect of the design of a suite of input models.  
\item
The design of the SKIRT {\tt{Geometry}} suite is based on a combination of basic building blocks and the extensive use of decorators. The building blocks can either be simple analytical components, or they can be numerical components defined as a set of smoothed particles or on a hierarchical or unstructured Voronoi grid. The {\tt{Geometry}} decorators combine and alter these building blocks to more complex structures.
\item
The different multivariate random number generation techniques that exist in the specialised numerical analysis literature can be used to efficiently implement complex decorators, for example those that add triaxiality, spiral structure or clumpiness to other models. Decorators can be chained without problems and essentially without limitation. Different layers of complexity can hence be added one by one. The result is that  very complex models can easily be constructed out of simple building blocks, without any coding at all.
\item
From the software design point of view, decorators have many advantages, including code transparency, the avoidance of code duplication, and an increase in code maintainability. This is a clear example that adhering to proven software design principles pays off, even for small and mid-sized projects.
\item
Finally, we demonstrate that our design using customised random position generators is superior to a simpler suite design based on a generic black-box random position generator. Using a suite of test simulations with increasing complexity we demonstrate that our customised random number generators are more accurate and more efficient.
\end{itemize}

\section*{Acknowledgements}

\noindent
This work fits in the CHARM framework (Contemporary physical challenges in Heliospheric and AstRophysical Models), a phase VII Interuniversity Attraction Pole (IAP) program organised by BELSPO, the BELgian federal Science Policy Office. The authors thank Ilse De Looze and S\'ebastien Viaene for their careful reading of a draft version of this paper, and all SKIRT users for their feedback, which has led to many improvements and additions to the code.

\end{document}